\documentclass[english, CRPHYS,Unicode,manuscript]{cedram}

\usepackage{graphicx}
\usepackage{dcolumn}
\usepackage{bm}
\usepackage{tikz}
\usepackage{pgfplots}
\usepackage{float}
\usepackage{mathrsfs}
\usepackage{amssymb}

\newcommand{\de}{\text{d}}

\newcommand{\dpar}[2]{\cfrac{\partial #1}{\partial#2}}

\newcommand{\vect}[1]{\mathbf{#1}}
\newcommand{\iac}[1]{{\color{blue} #1}}

\pgfplotsset{compat=1.17}

\title{Superradiant phononic emission from the analog spin ergoregion in a two-component Bose-Einstein condensate}

\author{\firstname{Anna} \lastname{Berti}}
\address{Pitaevskii BEC Center, CNR-INO and Dipartimento di Fisica, Università di Trento, I-38123 Trento, Italy}

\author{\firstname{Luca} \lastname{Giacomelli}}
\address{Université Paris Cité, CNRS, Matériaux et Phénomènes Quantiques, F-75013 Paris, France}
\address{Pitaevskii BEC Center, CNR-INO and Dipartimento di Fisica, Università di Trento, I-38123 Trento, Italy}

\author{\firstname{Iacopo} \lastname{Carusotto}}
\address{Pitaevskii BEC Center, CNR-INO and Dipartimento di Fisica, Università di Trento, I-38123 Trento, Italy}

\date{\today}

\begin{abstract}
	We make use of an analog gravity perspective to obtain a physical understanding of hydrodynamic instabilities stemming from the presence of quantized vortices in two-component atomic condensates and of their relation to ergoregion instabilities of rotating massive objects in gravitation. In addition to the localized instabilities related to vortex splitting, configurations displaying dynamically unstable modes that extend well outside the vortex core are found. In this case, the superradiant scattering process involves phonon emission into the much wider ergoregion of spin modes, so the physics most closely resembles the one of rotating massive objects. Our results confirm the potential of two-component condensates as analog models of rotating space-times in different regimes of gravitational interest.
\end{abstract}

\begin{document}

\maketitle

\section{Introduction}
Analog gravity \cite{Barcelo2005} is based on the use of table-top set-ups  
to simulate the physics of quantum fields on curved spacetimes. This approach is based on the formal equivalence of the equations of motion governing the dynamics of the two systems under suitable conditions. Since the original proposal by Unruh \cite{unruh1981}, curved space-time geometries have been realized in a large variety of systems, including surface waves on classical fluids \cite{rousseaux2008}, ultra-cold atoms \cite{lahav2010}, polariton fluids \cite{nguyen2015} and optical systems \cite{philbin2008, vocke2018}. 
The phenomena considered in this context typically involve strong gravity environments that escape direct astrophysical investigations, as in the vicinity of a black hole or during the fast expansion of the early universe. Most remarkable examples are Hawking emission from black hole horizons~\cite{belgiorno2010, weinfurtner2011, euve2016, steinhauer2016, munoz2019, drori2019}, cosmological particle creation and dynamical Casimir effect~\cite{wilson2011, jaskula2012,eckel2018,hung2013,wittemer2019,steinhauer2022analogue}, Penrose rotational superradiance \cite{torres2017, solnyshkov2019,cromb2020,braidotti2022measurement}. 

In this context, the term rotational superradiance  refers to the amplified scattering of waves from a rotating object at the expenses of the rotational energy of the object itself~\cite{Brito_2020}. It is a fully classical phenomenon and is predicted to occur in a variety of spacetime geometries, in particular in the surroundings of massive rotating objects. In this case, dragging of the spacetime by the rotating object gives rise to a so-called ergoregion, namely a spatial region supporting negative-energy light and matter modes.
The process of superradiance can be microscopically explained as the amplified reflection of a positive-energy wave impinging onto the boundary of the ergoregion, accompanied by the transmission of a negative-energy wave inside the ergoregion. In the presence of some reflecting element, self-amplification of one of the outgoing modes may take place, leading to an exponentially growing perturbation. Depending on the location of the reflecting element on either the inner or the outer side of the ergosurface, the instability takes the name of ergoregion instability~\cite{comins1978, friedman1978} or black hole-bomb instability~\cite{cardoso2004}.

Rotating Bose-Einstein condensates (BEC) are promising platforms to study superradiant phenomena~\cite{oliveira2018,Brito_2020}. 
A number of theoretical studies have investigated superradiance in configurations featuring a single vortex located at the center of a cylindrically symmetric trap as well as in planar geometries involving synthetic magnetic fields~\cite{giacomelli2021understanding}. In particular, the analog gravity perspective has offered transparent physical explanations for the hydrodynamic instability of multiply-charged single vortex configurations~\cite{fetter2001vortices, fetter2009rotating,fedichev1999dissipative, rokhsar1997vortex} 
by connecting them to ergoregion instabilities of rotating massive objects in gravitation~\cite{giacomelli2020ergoregion}. 

In the last decade, several authors have started investigating two-component BECs as a promising new platform for analog gravity~\cite{fischer2004quantum,visser2005massive,liberati2006analogue,butera2017black}. 
Such systems are characterized by the existence of two independent branches of collective excitations, associated to perturbations of the total density and of the spin density, featuring very different values of the speed of sound and of the healing length~\cite{cominotti2022observation}.
As a consequence of the much faster speed of density-sound, configurations displaying horizons for spin-sound waves can be engineered while keeping the superflow sub-sonic with respect to density-sound and therefore robust against perturbations coupling to the density. Furthermore, the direct experimental observation of the system dynamics is facilitated by the significantly larger size of the spin-healing length and by the availability of interferometric techniques based on a coherent mixing of the two components which allow to image all quadratures of the quantum field. On the long run, all this suggests the possibility of investigating quantum entanglement features of the superradiant  emission~\cite{finazzi2014entangled}.

Within this general framework, the goal of this paper is to explore ergoregion instability phenomena in two-component BECs displaying a single quantized vortex in both components. Thanks to the very different value of the density- and spin-sound speeds, the ergosurface for spin waves is no longer bound to sit inside or in the close vicinity of the vortex core as it instead happens in single component BECs, but can be pushed far away from the vortex core into the external region where the density is approximately constant. This is of great interest for analog gravity as it leaves a wider space to investigate the dynamics of the quantum field in the ergoregion and disentangle the different effects at play.

Beyond analog gravity, an active interest for this physics is also coming from a purely quantum gas perspective: a recent numerical work~\cite{kuopanportti2019} has pointed out the rich phenomenology of vortex splitting in harmonically trapped two-component BECs: singly-charged vortices are dynamically unstable for sufficiently strong repulsive inter-component interactions, whereas doubly-charged vortices dispose of several decay channels. Experimentally, the splitting of a singly-charged vortex into a pair of half-quantized vortices has been observed in spinor exciton-polariton superfluids \cite{manni2012dissociation} and antiferromagnetic spinor BECs \cite{seo2015half}.

Here, we take advantage of the analog gravity point of view to develop a transparent physical interpretation of vortex physics in two-component BECs in terms of superradiance effects. Use of this formalism allows us to highlight a number of features that are specific to two-components BECs and characterize them in view of forthcoming experimental studies. In addition to recovering the main predictions of previous works, e.g. on vortex splitting instabilities, we identify regimes where a clean analog of the ergoregion instability of massive rotating objects is visible. 

The structure of the paper is the following: in Secs.\ref{sec:bogo} and \ref{sec:vortices} we give a brief review of the theoretical tools that we are going to use in the following of the paper. In Sec. \ref{sec:lda} we tackle the vortex stability problem within the local density approximation (LDA) that provides a clear qualitative understanding of the physical mechanisms. A more quantitative analysis using a Bogoliubov characterization of the collective modes and a Gross-Pitaevskii (GP) simulation of the time-dependence is then provided in Secs.\ref{sec:spectra} and \ref{sec:GP}, respectively. A summary of our conclusions and of further extensions of our work is finally given in Sec.\ref{sec:concl}. Additional information on the numerical techniques adopted and on the independence of the results on the shape of the confinement potential are given in the Apppendices.

\section{Two-component condensates and Bogoliubov theory of density and spin collective excitations}
\label{sec:bogo}

In this work, we are interested in the dynamics of a two-component condensate subject to an external potential $V(r)$ and to a resonant coherent coupling of strength $\Omega\ge 0$ between the two components. Let us focus on a spin-symmetric configuration of equal masses $m_1=m_2\equiv m$ and intraspecies interactions $g_1 = g_2 \equiv g>0$. Moreover, we assume to deal with a miscible mixture, $|g_{12}|< g$, where $g_{12}$ defines the interspecies interaction strength.

The dynamics of such a system is ruled by two coupled GP equations \cite{stringari_2016}:
\begin{align}
\begin{split}
    i \hbar \partial_t \psi_1 &= \Big( K + V + g n_1 + g_{12} n_2  \Big) \psi_1 - \frac{\hbar \Omega}{2}\psi_{2} \\
    i \hbar\partial_t \psi_2 &= \Big(K + V + g n_2 + g_{12}n_1  \Big)\psi_2 - \frac{\hbar \Omega}{2} \psi_1
\end{split},
\label{eq:GP}
\end{align}
where $K = -\hbar^2 \nabla^2/2m $ is the kinetic energy operator, $\psi_j$ is the order parameter  of the $j$-th component, and $n_j = |\psi_j|^2$ is the associated atomic density, normalized to the number of particles in the $j$-th state, $N_j$. As long as $N_1 = N_2$, the ground state of the system is everywhere unpolarized, with identical order parameters $\psi_1^o(r)=\psi_2^o(r)$ and density profiles $n_1(r)=n_2(r)$ for the two components of the mixture. 

In addition to the $\mathbb Z_2$ spin-symmetry related to the exchange of the two components, for a vanishing coherent coupling $\Omega = 0$ the system has two additional $U(1)$ symmetries associated to the independent conservation of the number of particles $N_{1,2}$ in the two states. The presence of the coherent coupling allows particles to be transferred from one component to the other and therefore breaks one of these two $U(1)$ symmetries. In this case, we are left with a single $U(1)$ symmetry related to the conservation of the total atom number $N=N_1+N_2$ only.


A deep insight on the physical properties of the system is provided by the Bogoliubov theory of small perturbations of the collective excitations on top of the ground state. Thanks to the $\mathbb Z_2$ spin-symmetry of the problem under $\psi_1\leftrightarrow \psi_2$, excitations can be classified as symmetric, density ($d$) modes, and antisymmetric, spin ($s$) modes~\cite{Abad_2013}. 

\subsection{Uniform condensate}
For a uniform condensate of total density $n$, the energy of collective excitations  is given by the Bogoliubov dispersion relations as a function of the excitation wavevector $k$~\cite{Abad_2013}:
\begin{align}
    (\hbar \omega_d)^2 &=  \frac{\hbar^2 k^2}{2m} \left(\frac{\hbar^2 k^2}{2m} + 2\mu_d \right) \\
    (\hbar \omega_s)^2 &= \left(\frac{\hbar^2 k^2}{2m} + \hbar \Omega \right)\left(\frac{\hbar^2 k^2}{2m} + \hbar \Omega + 2\mu_s \right)\, ,\label{spin_disp}
\end{align}
where $\mu_d  = (g+g_{12})n/2$ and $\mu_s = (g-g_{12})n/2$ are the density and spin chemical potentials.

In the absence of Rabi coupling $\Omega=0$, both dispersions are linear at low momenta, with speed of sound equal to $c_d = \sqrt{\mu_d/m}$ and $c_s = \sqrt{\mu_s/m}$ in the density and spin channel, respectively; the presence of two gapless Goldstone modes is associated with the conservation of the total and relative numbers of particles, $N_1\pm N_2$. For a finite Rabi coupling $\Omega\ne 0$, instead, the breaking of the $U(1)$ symmetry related to the relative phase of the order parameters of the two components is responsible for the opening of a gap of size $\omega_p = \sqrt{\Omega (2\mu_s/\hbar + \Omega)}$ in the dispersion of spin modes, which acquire a finite effective mass~\cite{cominotti2022observation}. 

\subsection{Spatially inhomogeneous condensate}
For spin-symmetric, but spatially inhomogeneous systems, the Bogoliubov spectrum can only be obtained numerically (details are given in the Appendix): it is first neceassary to compute the density profile of a stationary state, that is, of an eigensolution of the stationary GP equation,
\begin{equation}
    \mu \Phi(\vect r) = \left[ K + V(\mathbf r) + (g+g_{12})\frac{n(\vect r)}{2} -\frac{\hbar\Omega}{2} \right] \Phi(\vect r)\,:
    \label{stat_GP}
\end{equation}
where $\Phi({\vect r})$ is the order parameter of both components, $\psi_1^o(\vect r)=\psi_2^o(\vect r)=\Phi(\vect r)$ and $\mu$ is the oscillation frequency of the matter field. Of course $\mu$ takes the meaning of chemical potential when we consider the lowest-order stationary state of \eqref{stat_GP} corresponding to the ground state of the condensate. In practice, the stationary state is obtained through imaginary-time evolution.

The frequencies $\omega_i$ and spatial shape $[u_i({\vect r}),v_i({\vect r})]$ of the collective excitation modes on top of the stationary state are then computed as the eigenvalues and eigenvectors of the Bogoliubov operators in the density $d$ and spin $s$ sectors:
\begin{align}
    \mathscr{L}_{d,s} &= \begin{pmatrix} 
 \mathcal D_{d,s} & \mu_{d,s}(\vect r) \\
    -\mu_{d,s}(\vect r) & - \mathcal D_{d,s}
    \end{pmatrix},   
    \label{Bogomat}
\end{align}
where the operators on the diagonal read:
\begin{align}
     \mathcal D_d &= K + V(\vect r) + 2\mu_d(\vect r) - \mu -\hbar \Omega/2\\
     \mathcal D_s &= K + V(\vect r) + \mu_s(\vect r) +\mu_d(\vect r) -\mu+\hbar \Omega/2 
\end{align}
and the interaction energies acquire a spatial dependence due to the inhomogeneity of the density profile, $\mu_{d,s}(\vect r)={(g\pm g_{12})}\,n({\vect r})/2$.
For each collective mode, the time-evolving perturbation of the order parameter of each species then has the form:
\begin{equation}
    \psi_i(\vect r,t) = \Phi(\vect r,t)+\delta\psi_i(\vect r,t)=
    \Phi(\vect r,t) + \Big[ u_i(\vect r) e^{-i\omega t}+ v_i^*(\vect r) e^{i\omega t} \Big] e^{-i\mu t/\hbar}\,, \label{bogoperturb}
\end{equation}
where the two components $u_i$ and $v_i$ of the eigenvector are often referred to as particle and antiparticle components. Depending on which of the two dominates over the other, excitation can have positive, negative or zero Bogoliubov norm:
\begin{equation}
    \|\delta\psi_i\|^2 = \int \sum_{i=1,2}\Big(|u_i(\vect r)|^2 - |v_i(\vect r)|^2 \Big) \de \vect r\,.
\end{equation}

In this work we exploit the analysis of the Bogoliubov spectrum~\cite{castin2001bose} to study the stability of stationary states, in particular vortex configurations. In stable configurations, all frequencies are real-valued and, for each excitation mode, have the same sign as the norm. 
Modes with positive frequency and negative norm (or vice versa) have instead a negative energy and their existence is a signature of an energetic (or thermodynamic) instability, whereas the configuration is dynamically unstable if some modes have a complex frequency. Thanks to the pseudo-hermiticity of the Bogoliubov matrix, these zero-norm modes always come in pairs, whose frequencies share the same real part and have opposite imaginary parts, $\omega = \eta \pm i\Gamma$. The imaginary part $\Gamma$ represents the rate of exponential growth of the dynamically unstable mode.

\section{Analog gravity and vortices in two-component BECs}
\label{sec:vortices}
Analog gravity is based on the formal equivalence between the Klein-Gordon equation for the dynamics of a scalar field on a curved spacetime and the linearized GP equations in the hydrodynamic regime in which all excitations propagate with the same speed of sound. In the particular case of excitations on top of a two-component BEC, the effective metric tensors of density and spin excitations are different, but share
the general form \cite{Barcelo2005}:
\begin{equation}
    (g_{d,s})_{\mu\nu} = \frac{n}{c_{d,s}}\begin{pmatrix} 
    -(c_{d,s}^2 - |\vect v|^2) & -\vect v^T \\ -\vect v & \mathbb I 
    \end{pmatrix}
    \label{analogmetric}
\end{equation}
where $\vect v(\vect r)$ is the (irrotational) velocity field associated to the background condensate flow:
\begin{equation}
    \vect v = \frac \hbar m \nabla \big( \text{arg}(\Phi) \big)\,.
\end{equation}
Let us now focus on the specific case of two-dimensional systems. Even though condensation is strictly speaking not possible in 2D, effectively two-dimensional atomic superfluids can be realized experimentally by tightly confining the condensate along one direction, say $z$, imposing a much weaker confinement in the orthogonal plane $(x,y)$, and working at a sufficiently low temperature for the coherence length to exceed the finite size of the system. The main effect of the vertical confinement is a geometrical renormalization of the interaction strength. 

In the following we focus our attention on a radially symmetric two-dimensional BEC confined in the $(x,y)$ plane. Thus $\vect r = (x,y)$ and polar coordinates are defined as $\theta = \arctan(y/x)$ and $r = \sqrt{x^2+y^2}$. For such configurations, the analog spacetime defined by \eqref{analogmetric} may display a density and a spin ergosurface, defined by $|\vect v| = c_{d,s}$, while analog event horizons are found if the radial fluid velocity equals the speed of sound, $|\vect v\cdot \vect u_r| = c_{d,s}$ and then exceeds it. Of course, a radial flow requires the presence of some particle non-conservation mechanism. A study of this physics goes beyond this paper, which is restricted to purely tangential flows with no radial component.

The ground state of a non-rotating condensate is well approximated by the Thomas-Fermi (TF) solution:
\begin{equation}
    \frac{n_\text{TF}(r)}{2} = |\Phi_\text{TF}(r)|^2 = \frac{\mu+\hbar\Omega/2 - V(r)}{g+g_{12}}\,.
\end{equation}
In the presence of rotation, a spin-symmetric quantized vortex is a solution of the stationary Gross-Pitaevski equation \eqref{stat_GP} of the form:
\begin{equation}
    \Phi(\vect r,t) = f(r) e^{iL\theta} e^{-i\mu t/\hbar}
    \label{eq:Phi_L}
\end{equation}
where the radial function $f(r)$ defines the vortex profile and $L$ is the integer-valued vortex charge. 
The density profile of a vortex differs from the TF one only in the region of the vortex core, while $n(r) \to n_\text{TF}(r)$ at larger distances. For a charge-$L$ vortex, the core has a size on the order of $\sqrt{L}\,\xi_d$, where $\xi_{d}=\sqrt{\hbar/mc_{d}}$ is the density healing length. The spin healing length is defined in an analogous way as $\xi_{s}=\sqrt{\hbar/mc_{s}}$.

As usual, the velocity flow of the vortex is tangential and inversely proportional to the distance $r$ from the center:
\begin{equation}
    \vect v(r) = \frac{\hbar L}{mr} \vect u_\theta 
\end{equation}
Assuming the two speeds of sound $c_{d,s}$ are slowly varying, the analog ergosurfaces for density and spin modes, defined by $|{\vect v}|=c_{d,s}$, are approximately located at $r\sim L\xi_{d,s}$. 

\begin{figure}[t]
    \centering
    \includegraphics[width = 0.55 \linewidth]{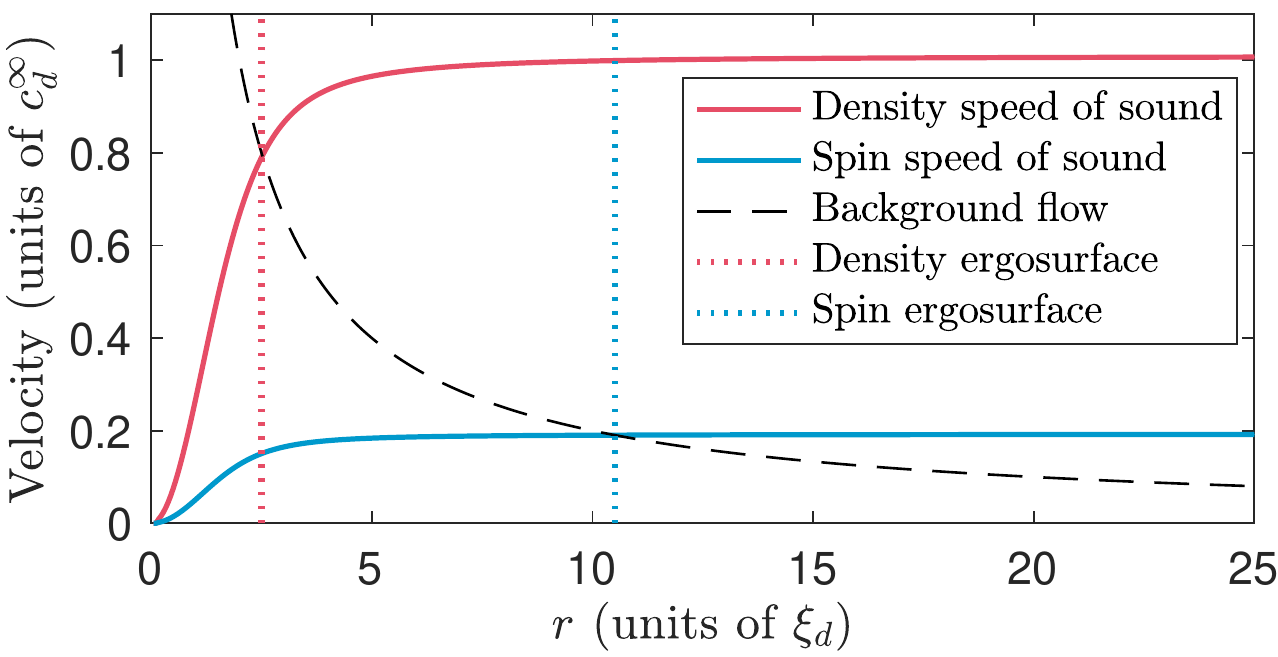}
    \caption{Comparison between the density and spin sound speed (red and blue line respectively), and the flow velocity (dashed black line) in a vortex configuration. The location of the ergosurface for density (spin) excitations, indicated by the dotted red (blue) line, is found as the intersection between $|\vect v(r)|$ and $c_d(r)$ [$c_s(r)$]. All velocities are normalized to the large-distance value of the speed of density sound, $c_d^\infty \equiv c_d(r\to\infty)$. 
    The plot is obtained using the numerically calculated density profile of a vortex of charge $L=2$; interactions are set to $g_{12}=0.93 g$, giving $c_d/c_s=\xi_s/\xi_d \sim 5.25$.   }
    \label{fig:speed}
\end{figure}

An example of promising configuration is shown in Fig.\ref{fig:speed}, where the local values of $c_{d,s}$ have been computed using the exact density profile of an $L=2$ vortex in a very wide trap and the interactions constants have been chosen such that the large-$r$ asymptotic values $c_s\ll c_d$ and $\xi_s \gg \xi_d$. This condition guarantees that the the spin ergoregion is located well outside the vortex core, where the density profile is approximately flat. 
The absence of a radial flow imposed by particle condensation prevents the system from having an analog event horizon, that would act as an absorbing boundary: the vortex is thus the ideal background to study analogs of the ergoregion instabilities of rotating space-times.

Taking advantage of the cylindrical symmetry, the Bogoliubov excitations of frequency $\omega$ and angular momentum $M$ on top of a charge-$L$ vortex can be written in the factorized form
\begin{align}
    u_i(\vect r) &= u_i(r) e^{iM\theta} e^{iL\theta}\\
    v_i(\vect r) &= v_i(r) e^{iM\theta} e^{-iL\theta}    
\end{align}
and the stability of the vortex configuration can be assessed by solving the effectively 1D radial Bogoliubov problem for $u_i(r)$ and $v_i(r)$. 
The Bogoliubov matrices \eqref{Bogomat} involve the kinetic energy operator. In polar coordinates this reads:
\begin{equation}
    K = -\frac{\hbar^2}{2m} \left( \dpar{^2}{r^2} + \frac{1}{r}\dpar{}{r} - \frac{(L\pm M)^2}{r^2} \right)
\end{equation}
where the $+$ ($-$) sign must be chosen when applying the operator to the particle (antiparticle) component $u$ ($v$).

\section{Local density approximation}
\label{sec:lda}

\begin{figure*}[t]
    \centering
    \includegraphics[width = \linewidth]{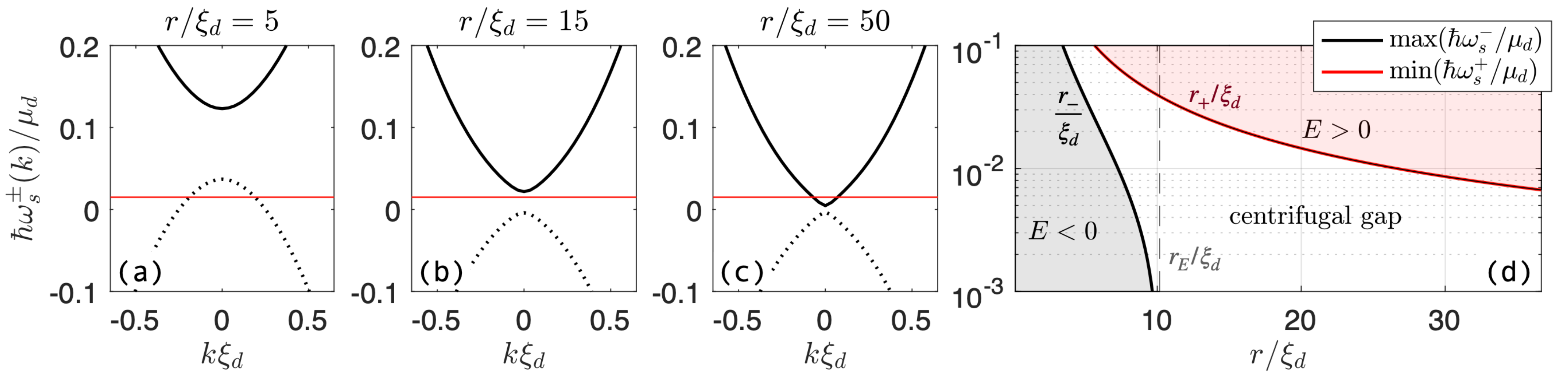}
    \caption{Properties of the dispersion relation in the LDA \eqref{bogodispM} (obtained with $L=2,M=1,g_{12}/g= 0.93,\Omega = 0$, $\mu_d/\mu_s = 27.5$). (a-c) Plot of the dispersion at different radii, showing the availability of positive and negative norm modes at a generic frequency $\omega$, indicated by the red horizontal line: solid (dotted) black lines refer to the upper (lower) branch. (d) Plot of the maximum frequency of the lower branch (black line) and minimum frequency of the upper branch (red line); the red (gray) area indicates the region of the plot where positive (negative) energy modes are available. The white area represents the centrifugal gap. The dashed black line indicates the spin ergosurface location, given by \eqref{ergoposition}.}
    \label{bogodisp}
\end{figure*}

In general, the Bogoliubov problem on top a spatially inhomogeneous vortex configuration can hardly be solved analytically. 
Before proceeding with the numerical diagonalization of the Bogoliubov matrices which will be the subject of the next Section, we start here with an intuitive discussion of the physics on the basis of a local density approximation (LDA).

Let us assume that the condensate is enclosed in a large box trap, so that the density can be considered approximately uniform far from the vortex core, $n(r\gg L\xi_d) \simeq n^\infty$. 
Within a small spatial region around $\vect r$, excitations of angular momentum $M$ can be approximated as plane waves with wavevector $\vect k = k\vect u_r + (M/r)\vect u_\theta$, so that the Bogoliubov wavefunctions $u(\vect r),v(\vect r) \propto e^{ikr}e^{iM\theta}$. By exploiting the LDA, the $k$-dependent Bogoliubov dispersion relation in the radial direction is obtained by taking the uniform result \eqref{spin_disp} and replacing $k^2 \to k^2+M^2/r^2$. The effect of the finite $M$ is thus to open a gap in the dispersion relation and is equivalent to having an additional space-dependent coherent coupling $\Omega_M(r) = \hbar M^2/2m r^2$.
On top of this, one needs to include the overall $M$-dependent Doppler shift $\delta(r)\equiv \vect k\cdot\vect v = \hbar LM/m r^2$ due to the tangential velocity flow. Due to the absence of in-going flow, this Doppler term does not tilt the dispersion relation as a function of the radial $k$, but only rigidly shifts it vertically along the frequency axis. This shift is crucial to bring negative-norm (positive-norm) modes up (down) to positive (negative) frequencies.

Combining all these elements together, we obtain the final formula
\begin{equation}
\omega_{s}^\pm (k) = \delta \pm \sqrt{\left(\frac{\hbar k^2}{2m} + \tilde\Omega\right) \left(\frac{\hbar k^2}{2m} + \tilde\Omega + \frac{2\mu_s^\infty}{\hbar} \right) }  
\label{bogodispM}
\end{equation}
where the sign $\pm$ refers to the positive and negative branches, respectively, and the effective chemical potential and coupling strength are $\mu_{s}^\infty  = (g-g_{12})n^\infty/2$ and $\tilde\Omega(r) = \Omega + \Omega_M(r)$. From this formula, the space-dependent centrifugal gap turns out to be $\Delta(r) \equiv \sqrt{ \tilde\Omega(r) \big[\tilde\Omega(r) +2\mu_s^\infty/\hbar\big] }$.

Because of the Doppler shift and the centrifugal gap, for a generic positive frequency $\omega$, there can be positive norm, negative norm or no modes available depending on the radial position $r$. Three specific examples for different radial positions $r$ are shown in Fig.\ref{bogodisp}(a-c) for the most relevant $\Omega=0$ case; including a finite $\Omega$ would simply result in a larger centrifugal gap. A summary of the results for generic $(r,\omega>0)$ is displayed in panel (d): here, the red and gray shaded areas indicate the $(r,\omega)$ regions where positive and negative norm modes are available, respectively. The boundaries of these regions identify two frequency-dependent positions in real space, $r_\pm(\omega)$, indicated by the red and black solid lines in Fig.\ref{bogodisp}(d). In particular, $r_-(\omega)$ can be thought as a sort of effective frequency-dependent ergosurface position. 

An ergoregion instability occurs when a negative norm mode in the inner part of the system is coupled to a positive norm mode living in the outer region, so that the two combine into a pair of zero-norm modes with complex frequencies $\omega \pm i\Gamma$~\cite{giacomelli2020ergoregion}. 
In other words, the dynamical instability can be thought as resulting from a tunnelling process between the red- and gray-shaded regions in Fig.\ref{bogodisp}(d), with a tunneling-mediated instability rate determined by the real-space width of the forbidden white region separating them. According to this interpretation, modes with smaller frequency, that is, hydrodynamic excitations, have a more extended antiparticle component, but also a smaller instability rate since they have to cross a much wider forbidden region. 

In general, a necessary (but not sufficient) requirement for the occurrence of an ergoregion instability is the presence of negative norm modes in the inner region, which translates in the condition $\delta(r)\ge \Delta(r)$ between the Doppler shift and the centrifugal gap. The limiting radius $r_{E}$ for which such condition is satisfied can be thought as the boundary of the ergoregion, that is the ergosurface.
In the absence of Rabi coupling ($\Omega=0$), the effective ergosurface for angular momentum $M$ modes is located at a radial position
\begin{equation}
    r_{E} = L\xi_s \sqrt{1-\frac{M^2}{4L^2}}\,. \label{ergoposition}
\end{equation}
This value is slightly reduced with respect to the hydrodynamic prediction $L\xi_s$, which is recovered in the $M\to 0$ limit, with the correction being due to the non-linear behaviour of the dispersion relation. For larger $M$ values, the ergoregion shrinks in space and eventually disappears for large angular momentum modes $M\ge 2L$. 

As mentioned above, the presence of a coherent coupling $\Omega \neq 0$ results in a larger centrifugal gap [white region in Fig.\ref{bogodisp}(d)], so the matching between the external positive-norm modes (red-shading) and the negative-norm ones in the ergoregion (gray-shading) modes  is harder to obtain and the unstable modes, if present, are more localized and display a weaker instability rate.

While this discussion based on the LDA provides an intuitive understanding of the microscopic process underlying the ergoregion instability, it completely misses all those features that stem from the quantization of the radial wavevector $k$ in a finite-size system\footnote{Differently from the planar configurations with synthetic vector potentials considered in~\cite{giacomelli2021understanding}, in our vortex configuration the irrotational nature of the superfluid flow fixes the characteristic spatial length on which the variation of ${\vect v}(r)$ can occur. As a result, it is not possible to find a rigorous limiting procedure in which the LDA is exactly verified.}. Since the radial wavevector is associated to the radial kinetic energy contribution $\hbar^2 k^2/2m$, we expect that for tight enough ergoregions, the minimum value of the kinetic energy set by the quantization of the negative norm mode may increase the size of the effective gap so much that it prevents the development of the instability. Assessing all these features in a quantitative way requires going beyond the LDA and solving the full Bogoliubov problem with numerical tools. This will be the subject of the next Section.



\section{Bogoliubov spectrum}
\label{sec:spectra}

The numerical solution of the Bogoliubov problem is performed as follows (more details can be found in the Appendix). For the chosen value of the vortex charge $L$, we first find the radial function $f(r)$ that describes the density profile of the stationary vortex and the associated oscillation frequency $\mu$. These are then used to build the Bogoliubov matrices for the density and spin excitations, whose diagonalization gives the eigenstates and the corresponding eigenvalues, that is, the Bogoliubov spectrum on top of the stationary state. 
In what follows we present results for harmonically trapped systems, but no qualitative difference is observed for box traps (see Appendix). In both cases we choose the parameters so that the radial size of the condensate $R$ is much larger than the vortex size. 

\begin{figure*}[t]
    \centering
    \includegraphics[width = \linewidth]{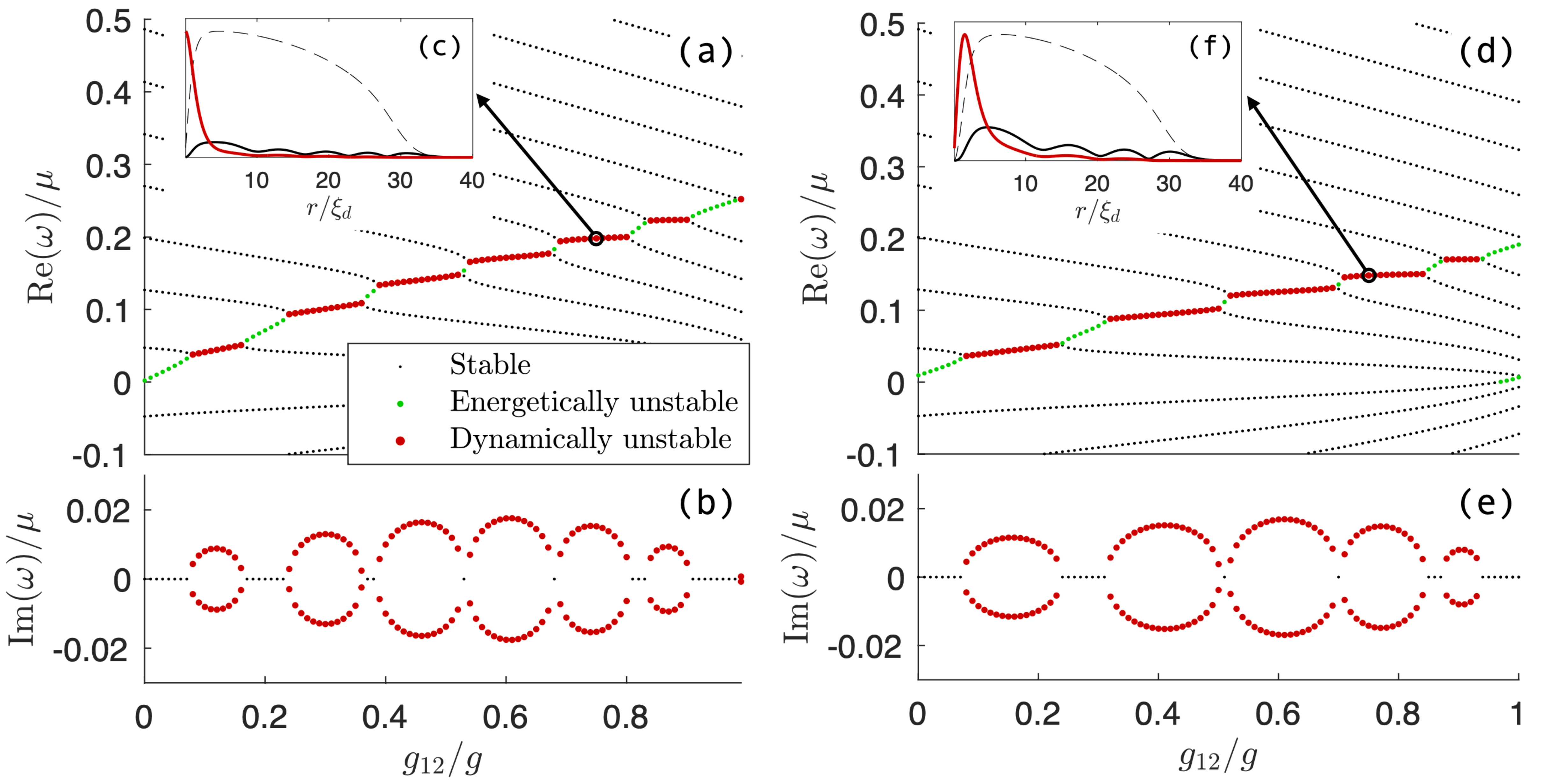}
    \caption{Bogoliubov spectrum for $M=1$ spin excitations on top of a $L=1$ (panels a-c) or $L=2$ (panels d-f) vortex located at the center of 
    a harmonically trapped mixture of TF radius $R=30\xi_d$. Panels (c) and (f) show, together with the radial profile of the BEC density (grey dashed line), an example of the real-space profile of the particle $|u(x)|$ (black solid line) and antiparticle $|v(x)|$ (red solid line) components of the dynamically unstable mode for $g_{12}=0.75 g$.
    }
    \label{fig:L1spectrum}
\end{figure*}

\begin{figure*}[t]
    \centering
    \includegraphics[width = \linewidth]{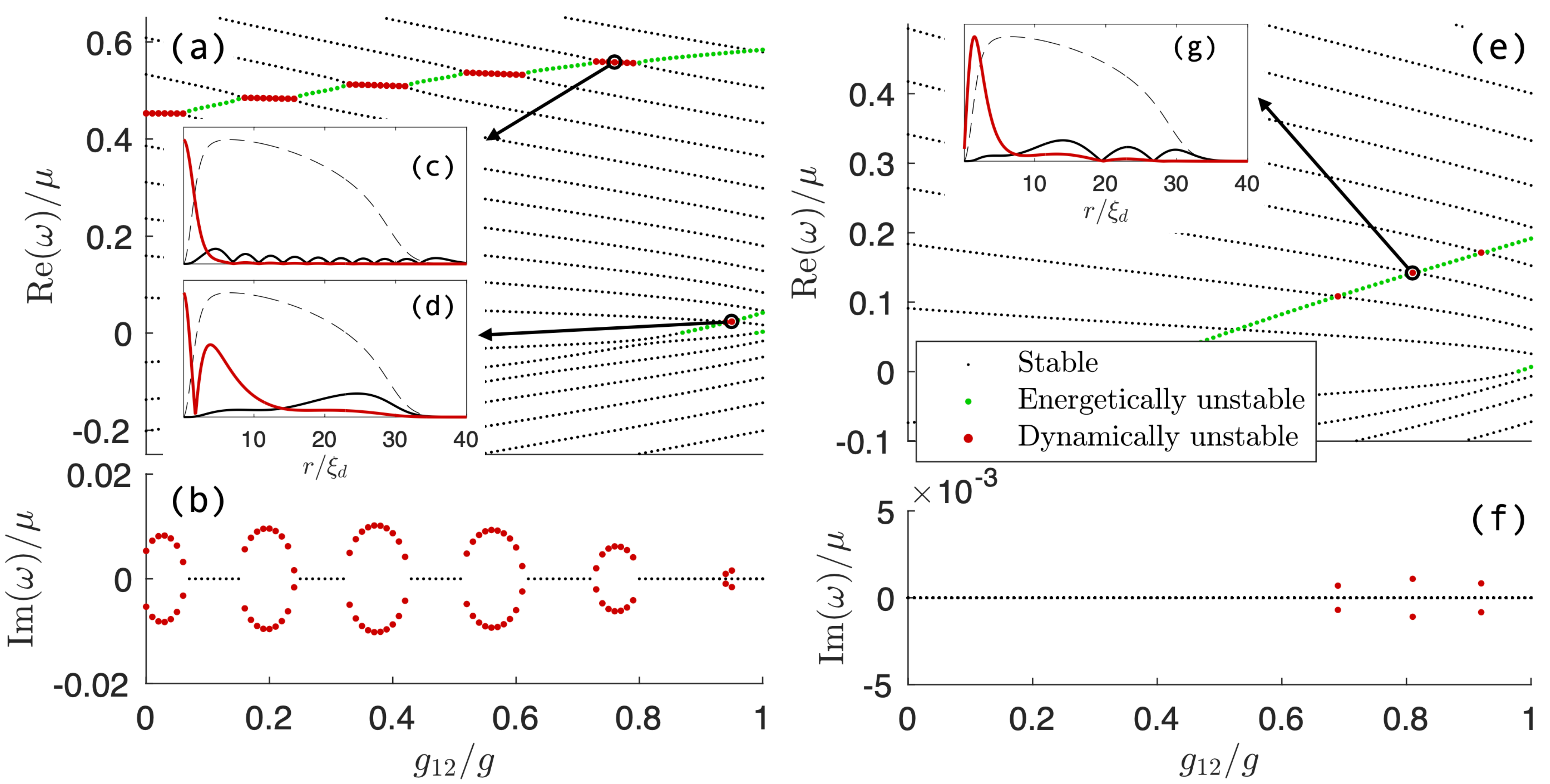}\caption{Bogoliubov spectrum for $M=2$ (panels a-d) and $M=3$ (panels e-g) spin excitations on top of an $L=2$ vortex in a harmonically trapped mixture of TF radius $R=30\xi_d$. 
    Two dynamical instabilities are present in the $M=2$ channel: the real-space profile of the particle (black solid line) and antiparticle (red solid line) components of these modes is shown in panels (c) and (d) for $g_{12}=0.75 g$ and $g_{12}=0.95$, respectively. The main difference between the two modes is in the localization of the antiparticle component, which dominates up to $r_- \sim 3\xi_d$ for the former, and up to $r_-\sim 14\xi_d$ for the latter; indeed, according to our LDA treatment, lower frequency modes are more extended.  Remarkably, the low frequency mode is only present for $g_{12}\sim g$. The $M=3$ channel shows a single instability: an example of the real-space profile of the mode is shown in panel (g) for $g_{12}=0.81 g$. 
    }
    \label{fig:L2spectrum}
\end{figure*}

Fig.\ref{fig:L1spectrum} and Fig.\ref{fig:L2spectrum} show the Bogoliubov spectra in the spin channel for different angular momenta $M$ as a function of the interaction strength ratio $g_{12}/g$ for vortices of charge $L=1,2$ in the $\Omega=0$ case: in correspondence of the crossing points between a positive and a negative energy mode, we observe the appearance of dynamically unstable modes (red dots) with the associated imaginary frequency bubble. The insets of Fig.\ref{fig:L1spectrum}(panels c, f) and Fig.\ref{fig:L2spectrum}(panels c,d,g) show examples of the spatial profiles of the particle and antiparticle components of the dynamically unstable modes: the common feature consists in the localization of the antiparticle component (red solid line) in the inner portion of the system, while the particle component (black solid line) is spread throughout the whole volume occupied by the fluid. 

In the case of singly-charged $L=1$ vortices, in the density channel a negative energy density mode is always present, but its frequency remains below that of positive norm modes (not shown); hence a dynamical instability never develops, in agreement with previous works~\cite{giacomelli2020ergoregion}. On the other hand, the spin channel shows energetic stability for attractive interspecies interactions $g_{12}<0$ (not shown) and alternated intervals of dynamical stability and instability for repulsive interactions $g_{12}>0$ [see Fig.\ref{fig:L1spectrum}(a-b)]. Similarly to what happens for density fluctuations of multiple-charged vortices in single-component BECs, this originates from the crossing of a spectrally isolated negative norm mode with the band of positive norm ones; the discreteness of the band, due to the finite size of the cloud, explains the existance of instability bubbles separated by intervals of dynamical stability\cite{giacomelli2020ergoregion}. Moreover, since the frequency of the isolated mode can be arbitrarily high, the associated instabilities are typically well localized in the vortex core [see Fig.\ref{bogodisp}(d)].   


As it is well known~\cite{fetter2001vortices}, when multiply-charged $L>1$ vortices are considered, dynamical instabilities of the same nature appear in the density channel too: since the energy of a vortex with charge $L$ is larger than that of $L$ singly-charged vortices, multiply charged vortices may be unstable against splitting into several lower-charge vortices. For instance, for $L=2$ vortices, we find, as a function of the interparticle interaction, alternate intervals of dynamical stability and instability in the $M=2$ channel of the density Bogoliubov problem (not shown), again in agreement with previous work~\cite{giacomelli2020ergoregion}.

An additional class of dynamical instabilities occur in the spin channels: besides the high-frequency isolated modes crossing the positive energy band discussed above, which are present for all angular momenta $1 \le M < 2L$ [see Fig.\ref{fig:L1spectrum}(d), Fig.\ref{fig:L2spectrum}(a,e)] and which are related to the deformation of the vortex core and/or its splitting, low-frequency and spatially extended instabilities may also appear. As visible in Fig.\ref{fig:L2spectrum}(a), such an instability may originate from the crossing of the two bands of closely spaced positive and negative norm modes that extend throughout the ergoregion.  As such it  belongs to the class of superradiant ergoregion instabilities closer to the hydrodynamic regime. The possibility of observing instabilities of this superradiant kind is one of the main interests of replacing single-component systems with spin mixtures with a much wider ergoregion.

Due to their different origin, these additional modes appear only around zero frequency for $g_{12}\lesssim g$ close to the demixing point: most importantly, their antiparticle component extends well outside the vortex core [see Fig.\ref{fig:L2spectrum}(d) as compared with Fig.\ref{fig:L2spectrum}(c)]. Physically, this means that long-wavelength spin waves are involved in the instability mechanism. From a quantum gas perspective, the instability can be understood as a demixing instability whose appearance is facilitated by the presence of a vortex which lowers its threshold. From the point of view of the gravitational analogy, this instability is of superradiant nature: as it happens for ergoregion instabilities around massive rotating objects, the positive energy of the wave in the outer region is compensated by the negative energy of the wave inside the ergoregion.

Analogously, in the presence of a coherent coupling, low-frequencies instabilities appear close to the ferromagnetic phase transition point $g_{12} \lesssim g + 2\Omega/n$, where the massive gap in the spin dispersion closes and the spin healing length diverges. This will be the subject of future work.

In general, depending on the parameters' values, multiple dynamically unstable modes can be found for the same configuration. When observing the time evolution of the system starting from an unperturbed $L$-charged vortex, after a short transient in which all the unstable modes compete, the mode with the largest imaginary frequency will eventually win over the others and dominate the intermediate-time evolution.  In most cases, the dominating mode is a high-frequency localized one. However, it is possible to play with the system size $R$ and with the interaction constants ratio $g_{12}/g$ to make all localized modes dynamically stable, while keeping a spatially extended mode unstable via superradiant mechanisms. In this case the dynamical instability is associated to the generation of long-wavelength phonon within the ergoregion, as expected from the gravitational analogy with ergoregion instabilities of massive rotating objects~\cite{Brito_2020}.

At very long times, nonlinear effects set in, resulting in complex mode mixing phenomena and additional instabilities. A study of this physics will be the subject of the next Section.


\section{Long-time dynamics}
\label{sec:GP}

It is well-known that the linearized Bogoliubov theory holds as long as the the excited modes are small perturbations on top of the stationary vortex state. When dealing with a dynamical instability, this approximation is valid for a limited amount of time, roughly given by the inverse of the growth rate $\Gamma$ of the unstable mode. After that, the cylindrical symmetry of the problem is likely to be broken by complex nonlinear processes, and mixing phenomena start occurring.

In order to access the long-time dynamics of the spin mixture, it is necessary to simulate the full GP equations \eqref{eq:GP} in two dimensions. The numerical protocol is the following (more details can be found in the Appendix): the stationary state corresponding to a vortex of charge $L$ sitting at $r=0$ is first obtained via a conjugate gradient algorithm \cite{antoine2017efficient}. In order to trigger the instability, we perturb the stationary state with some weak random noise, independently applied on the total and relative density. The temporal evolution of the system under the GPE \eqref{eq:GP} is then simulated using a split-step method.

\begin{figure*}[ht]
    \centering
    \includegraphics[width = \linewidth]{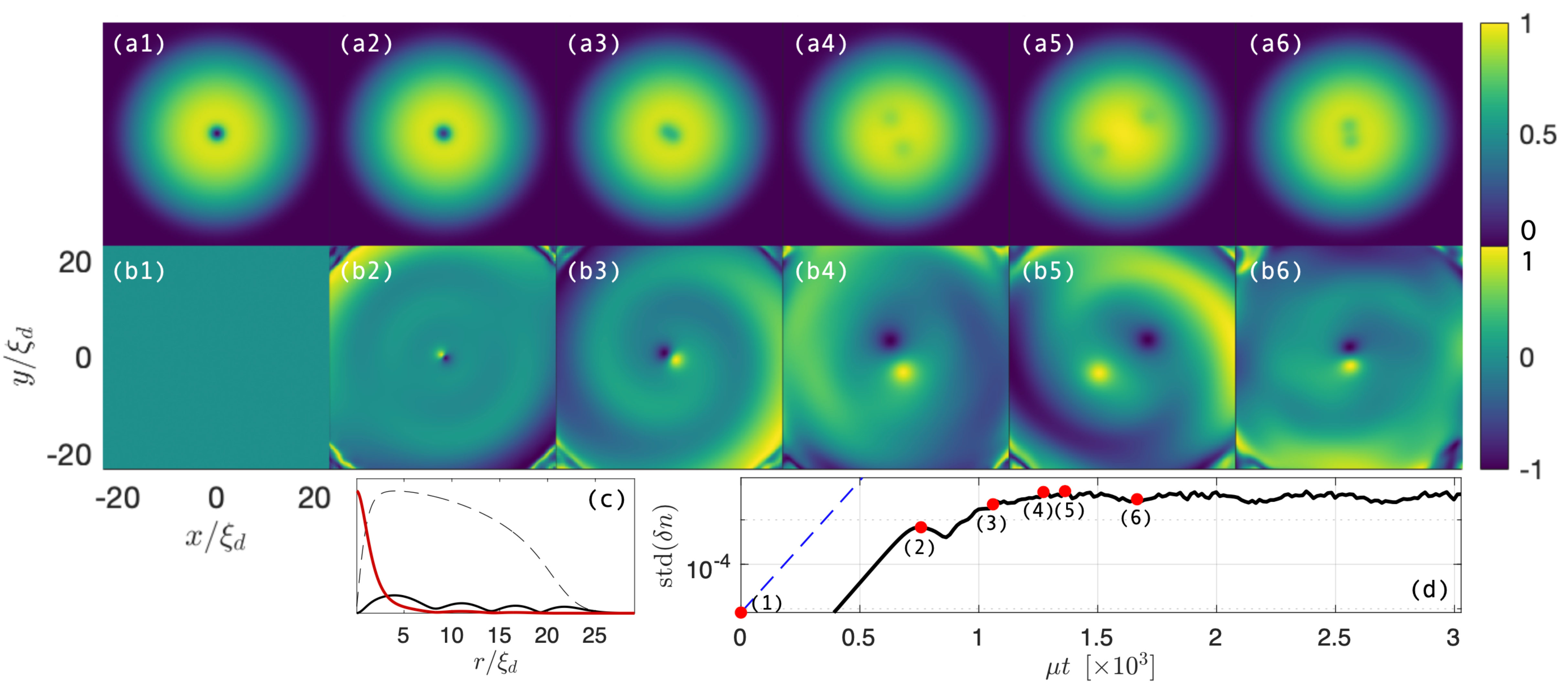}
    \caption{Dynamics of the vortex splitting instability.
    Real time evolution of a $L=1$ vortex with $R=21\xi_d$ and $g_{12}=0.8g$. For these parameters, the growth rate of the $M=1$ unstable mode, whose real-space profile is shown in panel (c), is $\Gamma/\mu \simeq 0.014$. Panels (a1-a6): total density $n = n_1+n_2$, measured in units of the peak Thomas-Fermi density $2 n_\text{TF}(r=0) = 2\mu/(g+g_{12})$. Panels (b1-b6): polarization, $Z = (n_1-n_2)/n$. Each column is computed at the time indicated by one of red dots in panel (d), which shows the temporal evolution of the standard deviation of the spin density $\delta n= n_1-n_2$ (black solid line), compared with the analytical exponential growth $\exp(\Gamma t)$ (blue dashed line).
    }
    \label{fig:L1splitting}
\end{figure*}

\begin{figure*}[ht]
    \centering
    \includegraphics[width = \linewidth]{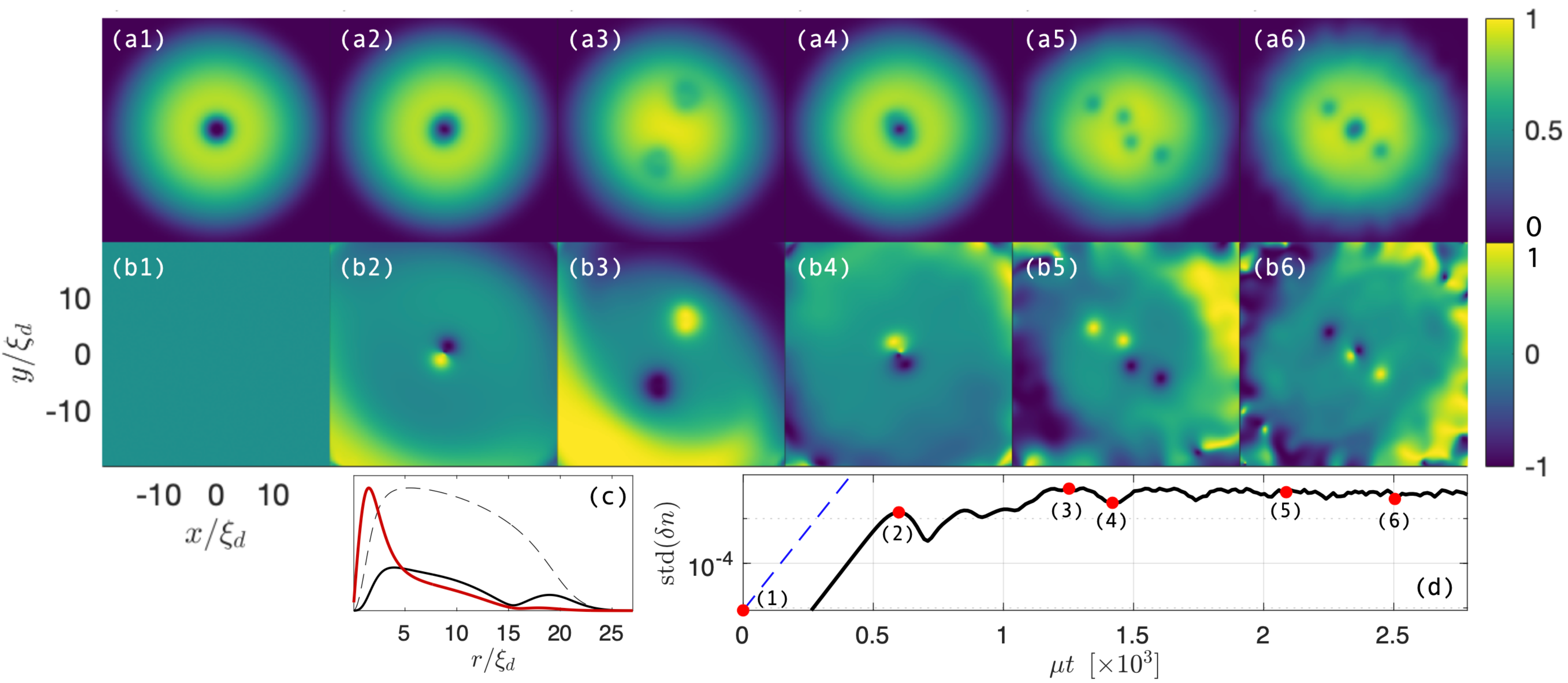}
    \caption{Dynamics of the vortex splitting instability. Same plots as in Fig.\ref{fig:L1splitting} for an $L=2$ vortex with $R= 19\xi_d$ and $g_{12}=0.42g$. For these parameters, the instability rate of the $M=1$ unstable mode, whose real-space profile is shown in panel (c), is $\Gamma/\mu = 0.019$. }
    \label{fig:L2splitting}
\end{figure*}

\begin{figure*}[t]
    \centering
    \includegraphics[width = \linewidth]{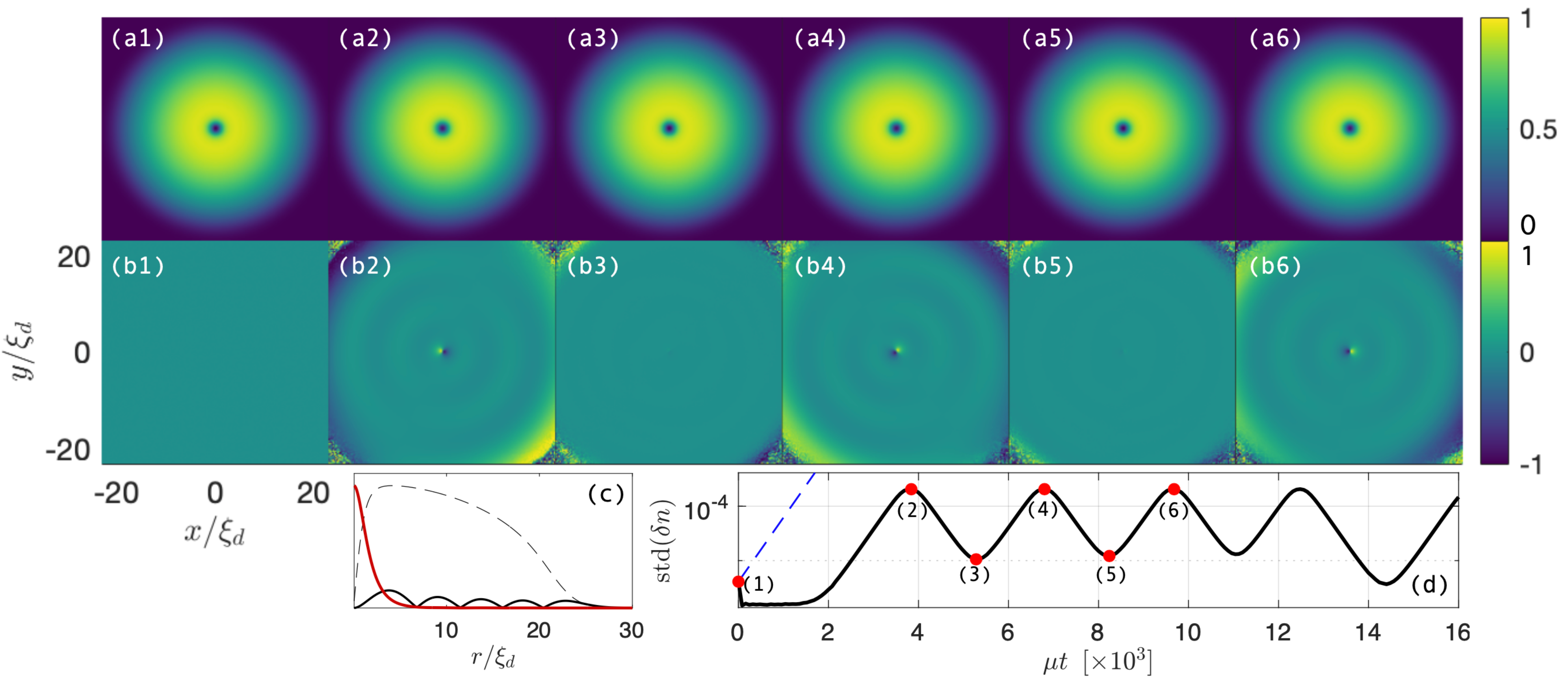}
    \caption{ Real time evolution of a $L=1$ vortex with $R=22\xi_d$ and $g_{12}=0.97g$. For these parameters, the growth rate of the $M=1$ unstable mode, whose real-space profile is shown in panel (c), is $\Gamma/\mu \simeq 0.0016$. Same plots as in Fig.\ref{fig:L1splitting}. }
  \label{fig:L1evol}
\end{figure*}

\begin{figure*}[ht]
    \centering
    \includegraphics[width = \linewidth]{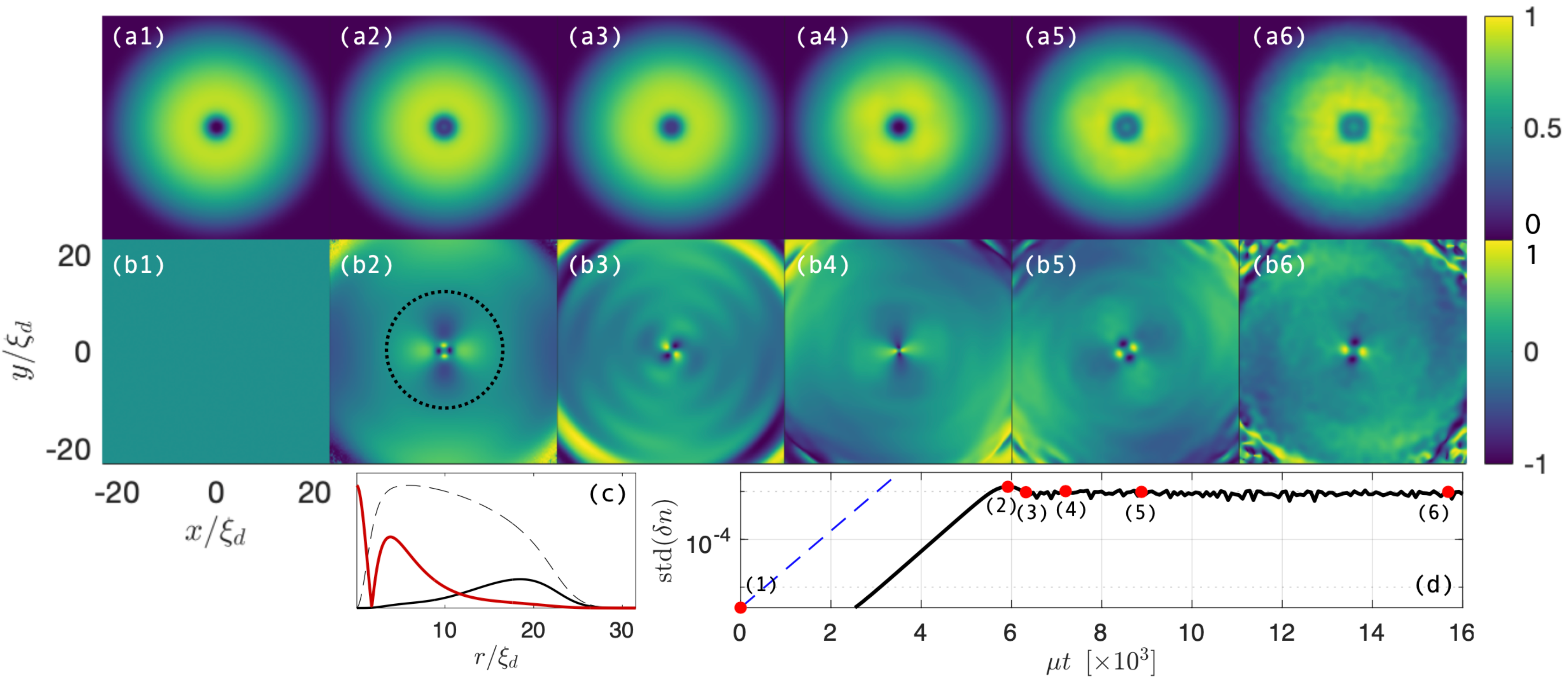}
    \caption{ Real time evolution of a $L=2$ vortex with $R=23\xi_d$ and $g_{12}=0.97g$. For these parameters, the growth rate of the $M=2$ unstable mode, whose real-space profile is shown in panel (c), is $\Gamma/\mu \simeq 0.0019$. Same plots as in Fig.\ref{fig:L1splitting}; the dotted black line in panel (b2) indicates the size of the region where the antiparticle component of the unstable mode dominates. }
  \label{fig:L2evol}
\end{figure*}

Examples of the simulated time-evolution are presented in Figs.\ref{fig:L1splitting}-\ref{fig:L2evol} for vortices of charge $L=1,2$. As already discussed, $L=1$ vortices have a single $M=1$ potentially unstable mode in the spin channel. 
On the other hand, $L=2$ vortices can feature up to five unstable modes, four of which in the spin channel (one for $M=1$, two for $M=2$, one for $M=3$). For the sake of simplicity, the simulation parameters are chosen in a way to have a single dynamical instability in the spin channel: in practice, this is done by keeping the interaction ratio $g_{12}/g$ fixed and tuning the system size $R$ to select the desired mode.

In all these figures, the short- and intermediate-time dynamics is consistent with our expectations based on the Bogoliubov theory: the instability develops initially as a well visible
perturbation characterized by $2M$ lobes in the azymuthal direction, where $M$ is the angular momentum of the unstable mode. For high-frequency instabilities, this excitation has a larger-amplitude part localized around $r=0$ and a smaller-amplitude one visible as interference fringes at larger radii [see panel (b2) of Figs.\ref{fig:L1splitting}-\ref{fig:L1evol}]; for low-frequencies instabilities instead, the localized part of the mode is itself spread over a larger portion of the system (see panel (b2) of Fig.\ref{fig:L2evol}).

In panel (d) of Figs.\ref{fig:L1splitting}-\ref{fig:L2evol}, we show the standard deviation (STD) of the spin density $\delta n = n_1-n_2$ from its ground state vanishing value, defined as 
\begin{equation}
\text{std}(\delta n) = \sqrt{\frac{1}{Q^2-1}\sum_{i,j=1}^Q \delta n^2(x_i, y_j) }
\end{equation}
where $Q\times Q$ is the number of points in the 2D numerical grid. 
As expected, this quantity is exponentially growing at a rate equal to the imaginary part $\Gamma$ of the frequency of the dynamically unstable mode, whereas the total density remains almost unperturbed. 

At longer times, $t\Gamma \gg 1$, for the intermediate values of $g_{12}/g$ considered in Figs.\ref{fig:L1splitting}-\ref{fig:L1evol}, the high-frequency dynamically unstable modes lead to the splitting of the original vortex into smaller vortices and/or pairs of half-quantized vortices. Two examples are shown in Figs.\ref{fig:L1splitting}-\ref{fig:L2splitting} for $L=1$ and $L=2$ vortices, respectively, and closely resemble the results of Ref.\cite{kuopanportti2019}. Interestingly, the splitting is often followed by a recombination process; similar dynamics was predicted for multiply-charged vortices  in a single-component BEC in \cite{patrick2022origin}. We attribute this phenomenon to energy conservation and to the finite size of the system: once the vortex is split, the excess of energy is released in the form of spin excitations, whose interference, after bouncing on the trap walls, reverses the splitting process. This typically occurs multiple times, before, in the absence of a dissipation mechanism, the evolution becomes turbulent, with additional vortices being nucleated from the boundaries and/or deformation of the (otherwise circular) cloud [see panels (a6-b6) of Fig.\ref{fig:L2splitting}].   
 All these complex nonlinear features go beyond this work and will be the subject of future studies.

The physics is more intriguing in the $g_{12}\lesssim g$ regime close to the demixing point, where
vortex splitting is much harder to observe and, if it does so, only occurs at much longer times scales: in addition to the reduced value of the instability rate $\Gamma$, the softness of spin modes prevents in fact the cloud from absorbing the excess energy  that would derive from vortex splitting. This result is confirmed by looking at the Bogoliubov spectra in Figs.\ref{fig:L1spectrum}-\ref{fig:L2spectrum}: for $g_{12}/g\lesssim 1$, the instability bubbles disappear or become tiny, and instability rates drop by roughly an order of magnitude. 
As a consequence, when approaching the demixing point, vortices in condensate mixtures remain stable for a very long time, while their spin dynamics is ruled by linear Bogoliubov theory. 

Two examples of the real-time dynamics in this regime are shown in Figs.\ref{fig:L1evol}-\ref{fig:L2evol} for $L=1,2$ vortices, respectively, and $g_{12}/g=0.97$. In the former case, the unstable Bogoliubov mode is spatially localized around the vortex and corresponds to a displacement of the cores in the two components,
with the consequent appearance of a net polarization. This process, however, does not lead to a full splitting of the vortex. We rather observe a sequence of intervals of suppression and revival of the instability, as witnessed by the oscillation in time of the standard deviation (STD) of the relative density. The underlying mechanism is rooted in the finite-size of the system and is analogous to the one leading to the sequence of vortex splitting and recombination events seen above.

The temporal evolution associated to a superradiant instability is finally visible in Fig.\ref{fig:L2evol}. At intermediate times [panel (b2)], a spatially extended $M=2$ modulation is clearly visible in association to the vortex core deformation. Its characteristic size corresponds to the spatial profile of the unstable mode shown in panel (c) and, in particular, its antiparticle component matches the extension of the ergoregion. This is the smoking gun of the superradiant nature of the instability.

At later times, the dynamics develops instead a more complicated behaviour that involves strong nonlinear mode-mixing effects: the perturbation of the spin density, due to the development of the instability, is large enough to excite the $M=2$ waves with higher radial momentum and several radial nodes that are visible in panel (b3). These different excitation patterns are in competition, eventually leading to a marked deformation of the vortex and a significant modulation also of the total density [see panel (a6)]. Once again, the accurate analysis of these complex mode-mixing processes goes beyond the scope of the present work.



\section{Conclusions and perspectives}
\label{sec:concl}
In this work, we have theoretically analysed the stability of quantized vortices in symmetric two-component atomic Bose-Einstein condensates (BEC) and we have interpreted the results within an analog gravity context in terms of ergoregion instabilities. In addition to instabilities related to high-frequency Bogoliubov modes localized in the vortex core and associated to the distortion or the splitting of the vortex core  like in single-component BECs~\cite{giacomelli2020ergoregion}, suitable regimes are found where the physics is determined by low-frequency and long-wavelength spin excitations. In this case, the superradiant scattering process underlying the ergoregion instability involves the excitation of a Doppler-shifted, negative-energy spin-sound wave spread over the ergoregion and the simultaneous emission of positive-energy spin waves into the outer part of the BEC, in close analogy with the ergoregion instability of space-time around a rotating massive object.

On the theoretical side, the most challenging open questions concern the late time dynamics of unstable configurations. First studies of this physics focused on the simpler case of black hole laser instabilities~\cite{deNova:PRA2016} and first hints of the remarkable complexity of the superradiant case are visible in the simulations reported in this work. In this long-term adventure, a special attention has to be paid to the role of nonlinear processes in saturating the instability, in analogy with related phenomena predicted in the gravitational context such as the growth of the so-called black hole ``hair''~\cite{Bos16,San16,Eas17,Ike19}. 

On the experimental side, two-component condensates can be obtained in fluids of atoms and of light. In the atomic case, vortices can be generated by means of suitably chosen stirring potential~\cite{Madison:PRL2000,fetter2009rotating} and the different components can be chosen within the atomic hyperfine structure in a way to obtain slow spin-sound waves~\cite{cominotti2022observation}. 
In the optical case, the polarization degree of freedom can be used to obtain two-component condensates and arbitrary velocity patterns can be imprinted onto the fluid by suitably designing the phase profile of the pump beam so to generate rotating configurations~\cite{Carusotto:RMP2013}. 
In both cases, the control of the coherent coupling strength via either an external electromagnetic dressing of the atoms or the optical birefringence of the cavity material may be used to tune the mass of the quantum field. 

The remarkable experimental advances in both fields make us confident that it will be soon possible to validate our conclusions using state-of-the-art set-ups. 
As a more ambitious challenge, the analysis of correlations in the spirit of \cite{carusotto2008numerical,recati2009bogoliubov,steinhauer2016}, will allow to investigate superradiant processes at the quantum level, so to prove superradiant amplification is intrinsically connected to the spontaneous creation of correlated pairs of Bogoliubov modes with opposite energy at the ergosurface.


\section{Acknowledgements}
AB aknowledges financial support from Q@TN, the joint lab between University of Trento, FBK- Fondazione Bruno Kessler, INFN- National Institute for Nuclear Physics and CNR- National Research Council.
IC acknowledges financial support from the European Union H2020-FETFLAG-2018-2020 project ``PhoQuS'' (n.820392), from the Provincia Autonoma di Trento, and from the Q@TN initiative.

\section{Appendix}

\subsection{Numerical details}
The numerical calculation of the Bogoliubov spectrum is performed as follows. Given a value of the vortex charge $L$, we first find the radial function $f(r)$ that describes the density profile of the stationary vortex and the associated oscillation frequency $\mu$. This is achieved by means of an imaginary-time evolution . 
Once the vortex profile $f(r)$ is known, for each value of angular momentum $1\leq M<2L$, we build the density and spin Bogoliubov matrices and diagonalize them: the eigenvalues form the Bogoliubov spectrum of the vortex, while the eigenvectors give the real-space profiles of the modes. 

The free parameters of the simulation are the interaction ratio $g_{12}/g$, the Rabi frequency $\Omega$, the radial size of the system $R$ and the expected chemical potential in the TF regime (i.e. without vortices) $\mu_{TF}$. The remaining parameters, i.e. the number of particles $N$ and the trapping frequency $\omega_0$, are chosen accordingly; in the case of a uniform mixture in a box of radius $R$, we set:
\begin{equation}
    \frac{N}{\pi R^2} =   \frac{2\mu_{TF} + \hbar\Omega}{g+g_{12}} 
\end{equation}
while, for a system in an harmonic trap of TF radius $R$, we choose:
\begin{equation}
    \omega_0 R = \sqrt{ \frac{2\mu_{TF} + \hbar\Omega}{m}} \qquad \textrm{and} \qquad \frac{2N}{\pi R^2} =   \frac{2\mu_{TF} + \hbar\Omega}{g+g_{12}} 
\end{equation}

While we can take advantage of the axial symmetry of the system to compute the Bogoliubov spectra, the same strategy is not applicable to the problem of determining the long-time dynamics of the mixture: we expect indeed the cylindrical symmetry to be broken by the development of the instability. 

We thus solve the full GP equations in 2D, according to the following numerical protocol: we first identify, thanks to the Bogoliubov spectra, a set of parameters leading to a dynamical instability; we calculate the 2D profile of the stationary vortex via a conjugate gradient algorithm \cite{antoine2017efficient}, where the azymuthal profile of the wavefunction phase is enforced at every step. 
The stationary state is perturbed with some weak random noise, applied independently on the total and relative density, to trigger the instability. We then simulate the time evolution of the system given by the GPE \eqref{eq:GP} using a split-step method.

\subsection{Dependence on the size and on the external potential}
\begin{figure}
    \centering
    \includegraphics[width = \linewidth]{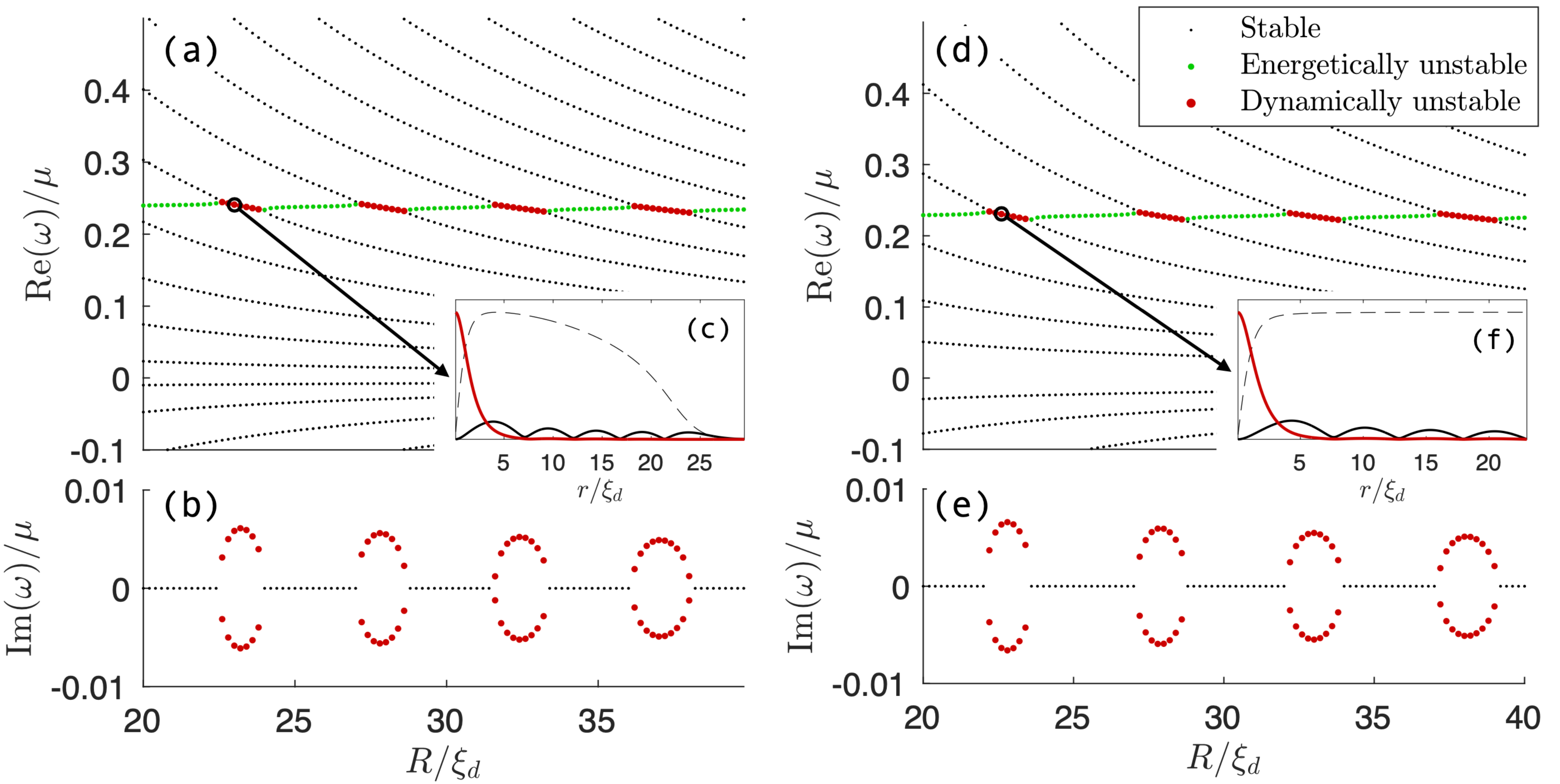}
    \caption{Bogoliubov spectra for $M=1$ spin excitations over a vortex of charge $L=1$ with $g_{12}=0.93g$. Panels (a-c) refer to an harmonically trapped system of TF radius $R$, while panels (d-f) refer to a uniform mixture in a box of size $R$. Panels (c) and (f) show the real-space profile of the particle (black solid line) and antiparticle (red solid line) components of the dynamically unstable mode for $R=23\xi_d$; the dashed line represents the rescaled vortex profile $f(r)$.  }
    \label{fig:appendix1}
\end{figure}

In order to assess the dependence of dynamical instabilities on the size of the system and on the external potential applied to the atoms, we performed a calculation of the Bogoliubov spectra as a function of $R$ for two experimentally relevant cases: the harmonic trap and the box. 
The results are shown in Fig.\ref{fig:appendix1} for $L=M=1$: in agreement with Ref.\cite{giacomelli2020ergoregion}, for both cases we find alternate intervals of dynamical stability and instability; as shown in panels (b) and (e), the instability bubbles acquire a larger width but smaller amplitude as $R$ increases. 
This is the reason why in the main text, if possible, we simulate the real-time dynamics of systems with relatively small size ($R\sim 20\xi_d$): in addition to the advantage of allowing for a better resolution in space, such configurations also feature a faster development of the instabilities. As a further check, panels (c) and (f) of Fig.\ref{fig:appendix1} show that there is no qualitative difference in the real-space profiles of the unstable modes given \iac{in} the two examined configurations. We verified that this holds true also for other values of vortex charge $L$ and angular momentum $M$ (not shown). 

\bibliographystyle{crunsrt}

\nocite{*}
\bibliography{biblio}

\providecommand{\noopsort}[1]{}\providecommand{\singleletter}[1]{#1}%
\def\bysame{\leavevmode ---------\thinspace}
\makeatletter\if@francais\providecommand{\og}{<<~}\providecommand{\fg}{~>>}
\else\gdef\og{``}\gdef\fg{''}\fi\makeatother
\def\cdrandname{\&}
\providecommand\cdrnumero{no.~}
\providecommand{\cdredsname}{eds.}
\providecommand{\cdredname}{ed.}
\providecommand{\cdrchapname}{chap.}
\providecommand{\cdrmastersthesisname}{Memoir}
\providecommand{\cdrphdthesisname}{PhD Thesis}
\begin{thebibliography}{10}

\bibitem{Barcelo2005}
C.~Barceló, S.~Liberati, M.~Visser, {\og Analogue Gravity\fg}, \emph{Living
  Reviews in Relativity} \textbf{8} (2005), \cdrnumero 1,
  \url{http://dx.doi.org/10.12942/lrr-2005-12}.

\bibitem{unruh1981}
W.~G. Unruh, {\og Experimental black-hole evaporation?\fg}, \emph{Physical
  Review Letters} \textbf{46} (1981), \cdrnumero 21, p.~1351.

\bibitem{rousseaux2008}
G.~Rousseaux, C.~Mathis, P.~Ma{\"\i}ssa, T.~G. Philbin, U.~Leonhardt, {\og
  Observation of negative-frequency waves in a water tank: a classical analogue
  to the Hawking effect?\fg}, \emph{New Journal of Physics} \textbf{10} (2008),
  \cdrnumero 5, p.~053015.

\bibitem{lahav2010}
O.~Lahav, A.~Itah, A.~Blumkin, C.~Gordon, S.~Rinott, A.~Zayats, J.~Steinhauer,
  {\og Realization of a sonic black hole analog in a Bose-Einstein
  condensate\fg}, \emph{Physical review letters} \textbf{105} (2010),
  \cdrnumero 24, p.~240401.

\bibitem{nguyen2015}
H.~S. Nguyen, D.~Gerace, I.~Carusotto, D.~Sanvitto, E.~Galopin,
  A.~Lema{\^\i}tre, I.~Sagnes, J.~Bloch, A.~Amo, {\og Acoustic black hole in a
  stationary hydrodynamic flow of microcavity polaritons\fg}, \emph{Physical
  review letters} \textbf{114} (2015), \cdrnumero 3, p.~036402.

\bibitem{philbin2008}
T.~G. Philbin, C.~Kuklewicz, S.~Robertson, S.~Hill, F.~Konig, U.~Leonhardt,
  {\og Fiber-optical analog of the event horizon\fg}, \emph{Science}
  \textbf{319} (2008), \cdrnumero 5868, p.~1367-1370.

\bibitem{vocke2018}
D.~Vocke, C.~Maitland, A.~Prain, K.~E. Wilson, F.~Biancalana, E.~M. Wright,
  F.~Marino, D.~Faccio, {\og Rotating black hole geometries in a
  two-dimensional photon superfluid\fg}, \emph{Optica} \textbf{5} (2018),
  \cdrnumero 9, p.~1099-1103.

\bibitem{belgiorno2010}
F.~Belgiorno, S.~L. Cacciatori, M.~Clerici, V.~Gorini, G.~Ortenzi, L.~Rizzi,
  E.~Rubino, V.~G. Sala, D.~Faccio, {\og Hawking radiation from ultrashort
  laser pulse filaments\fg}, \emph{Physical review letters} \textbf{105}
  (2010), \cdrnumero 20, p.~203901.

\bibitem{weinfurtner2011}
S.~Weinfurtner, E.~W. Tedford, M.~C. Penrice, W.~G. Unruh, G.~A. Lawrence, {\og
  Measurement of stimulated Hawking emission in an analogue system\fg},
  \emph{Physical review letters} \textbf{106} (2011), \cdrnumero 2, p.~021302.

\bibitem{euve2016}
L.-P. Euv{\'e}, F.~Michel, R.~Parentani, T.~G. Philbin, G.~Rousseaux, {\og
  Observation of noise correlated by the Hawking effect in a water tank\fg},
  \emph{Physical review letters} \textbf{117} (2016), \cdrnumero 12, p.~121301.

\bibitem{steinhauer2016}
J.~Steinhauer, {\og Observation of quantum Hawking radiation and its
  entanglement in an analogue black hole\fg}, \emph{Nature Physics} \textbf{12}
  (2016), \cdrnumero 10, p.~959-965.

\bibitem{munoz2019}
J.~R. Mu{\~n}oz~de Nova, K.~Golubkov, V.~I. Kolobov, J.~Steinhauer, {\og
  Observation of thermal Hawking radiation and its temperature in an analogue
  black hole\fg}, \emph{Nature} \textbf{569} (2019), \cdrnumero 7758,
  p.~688-691.

\bibitem{drori2019}
J.~Drori, Y.~Rosenberg, D.~Bermudez, Y.~Silberberg, U.~Leonhardt, {\og
  Observation of stimulated Hawking radiation in an optical analogue\fg},
  \emph{Physical Review Letters} \textbf{122} (2019), \cdrnumero 1, p.~010404.

\bibitem{wilson2011}
C.~M. Wilson, G.~Johansson, A.~Pourkabirian, M.~Simoen, J.~R. Johansson,
  T.~Duty, F.~Nori, P.~Delsing, {\og Observation of the dynamical Casimir
  effect in a superconducting circuit\fg}, \emph{nature} \textbf{479} (2011),
  \cdrnumero 7373, p.~376-379.

\bibitem{jaskula2012}
J.-C. Jaskula, G.~B. Partridge, M.~Bonneau, R.~Lopes, J.~Ruaudel, D.~Boiron,
  C.~I. Westbrook, {\og Acoustic analog to the dynamical Casimir effect in a
  Bose-Einstein condensate\fg}, \emph{Physical Review Letters} \textbf{109}
  (2012), \cdrnumero 22, p.~220401.

\bibitem{eckel2018}
S.~Eckel, A.~Kumar, T.~Jacobson, I.~B. Spielman, G.~K. Campbell, {\og A rapidly
  expanding Bose-Einstein condensate: an expanding universe in the lab\fg},
  \emph{Physical Review X} \textbf{8} (2018), \cdrnumero 2, p.~021021.

\bibitem{hung2013}
C.-L. Hung, V.~Gurarie, C.~Chin, {\og From cosmology to cold atoms: observation
  of Sakharov oscillations in a quenched atomic superfluid\fg}, \emph{Science}
  \textbf{341} (2013), \cdrnumero 6151, p.~1213-1215.

\bibitem{wittemer2019}
M.~Wittemer, F.~Hakelberg, P.~Kiefer, J.-P. Schr{\"o}der, C.~Fey,
  R.~Sch{\"u}tzhold, U.~Warring, T.~Schaetz, {\og Phonon pair creation by
  inflating quantum fluctuations in an ion trap\fg}, \emph{Physical review
  letters} \textbf{123} (2019), \cdrnumero 18, p.~180502.

\bibitem{steinhauer2022analogue}
J.~Steinhauer, M.~Abuzarli, T.~Aladjidi, T.~Bienaim{\'e}, C.~Piekarski, W.~Liu,
  E.~Giacobino, A.~Bramati, Q.~Glorieux, {\og Analogue cosmological particle
  creation in an ultracold quantum fluid of light\fg}, \emph{Nature
  Communications} \textbf{13} (2022), \cdrnumero 1, p.~1-7.

\bibitem{torres2017}
T.~Torres, S.~Patrick, A.~Coutant, M.~Richartz, E.~W. Tedford, S.~Weinfurtner,
  {\og Rotational superradiant scattering in a vortex flow\fg}, \emph{Nature
  Physics} \textbf{13} (2017), \cdrnumero 9, p.~833-836.

\bibitem{solnyshkov2019}
D.~Solnyshkov, C.~Leblanc, S.~Koniakhin, O.~Bleu, G.~Malpuech, {\og Quantum
  analogue of a Kerr black hole and the Penrose effect in a Bose-Einstein
  condensate\fg}, \emph{Physical Review B} \textbf{99} (2019), \cdrnumero 21,
  p.~214511.

\bibitem{cromb2020}
M.~Cromb, G.~M. Gibson, E.~Toninelli, M.~J. Padgett, E.~M. Wright, D.~Faccio,
  {\og Amplification of waves from a rotating body\fg}, \emph{Nature Physics}
  \textbf{16} (2020), \cdrnumero 10, p.~1069-1073.

\bibitem{braidotti2022measurement}
M.~C. Braidotti, R.~Prizia, C.~Maitland, F.~Marino, A.~Prain, I.~Starshynov,
  N.~Westerberg, E.~M. Wright, D.~Faccio, {\og Measurement of penrose
  superradiance in a photon superfluid\fg}, \emph{Physical Review Letters}
  \textbf{128} (2022), \cdrnumero 1, p.~013901.

\bibitem{Brito_2020}
R.~Brito, V.~Cardoso, P.~Pani, \emph{Lecture Notes in Physics} (2020),
  \url{http://dx.doi.org/10.1007/978-3-030-46622-0}.

\bibitem{comins1978}
N.~Comins, B.~F. Schutz, {\og On the ergoregion instability\fg},
  \emph{Proceedings of the Royal Society of London. A. Mathematical and
  Physical Sciences} \textbf{364} (1978), \cdrnumero 1717, p.~211-226.

\bibitem{friedman1978}
J.~L. Friedman, {\og Ergosphere instability\fg}, \emph{Communications in
  Mathematical Physics} \textbf{63} (1978), \cdrnumero 3, p.~243-255.

\bibitem{cardoso2004}
V.~Cardoso, O.~J.~C. Dias, J.~P.~S. Lemos, S.~Yoshida, {\og Black-hole bomb and
  superradiant instabilities\fg}, \emph{Phys. Rev. D} \textbf{70} (2004),
  p.~044039, \url{https://link.aps.org/doi/10.1103/PhysRevD.70.044039}.

\bibitem{oliveira2018}
L.~A. Oliveira, L.~J. Garay, L.~C.~B. Crispino, {\og Ergoregion instability of
  a rotating quantum system\fg}, \emph{Phys. Rev. D} \textbf{97} (2018),
  p.~124063, \url{https://link.aps.org/doi/10.1103/PhysRevD.97.124063}.

\bibitem{giacomelli2021understanding}
L.~Giacomelli, I.~Carusotto, {\og Understanding superradiant phenomena with
  synthetic vector potentials in atomic Bose-Einstein condensates\fg},
  \emph{Physical Review A} \textbf{103} (2021), \cdrnumero 4, p.~043309.

\bibitem{fetter2001vortices}
A.~L. Fetter, A.~A. Svidzinsky, {\og Vortices in a trapped dilute Bose-Einstein
  condensate\fg}, \emph{Journal of Physics: Condensed Matter} \textbf{13}
  (2001), \cdrnumero 12, p.~R135.

\bibitem{fetter2009rotating}
A.~L. Fetter, {\og Rotating trapped bose-einstein condensates\fg},
  \emph{Reviews of Modern Physics} \textbf{81} (2009), \cdrnumero 2, p.~647.

\bibitem{fedichev1999dissipative}
P.~Fedichev, G.~Shlyapnikov, {\og Dissipative dynamics of a vortex state in a
  trapped Bose-condensed gas\fg}, \emph{Physical Review A} \textbf{60} (1999),
  \cdrnumero 3, p.~R1779.

\bibitem{rokhsar1997vortex}
D.~Rokhsar, {\og Vortex stability and persistent currents in trapped Bose
  gases\fg}, \emph{Physical review letters} \textbf{79} (1997), \cdrnumero 12,
  p.~2164.

\bibitem{giacomelli2020ergoregion}
L.~Giacomelli, I.~Carusotto, {\og Ergoregion instabilities in rotating
  two-dimensional Bose-Einstein condensates: Perspectives on the stability of
  quantized vortices\fg}, \emph{Physical Review Research} \textbf{2} (2020),
  \cdrnumero 3, p.~033139.

\bibitem{fischer2004quantum}
U.~R. Fischer, R.~Sch{\"u}tzhold, {\og Quantum simulation of cosmic inflation
  in two-component Bose-Einstein condensates\fg}, \emph{Physical Review A}
  \textbf{70} (2004), \cdrnumero 6, p.~063615.

\bibitem{visser2005massive}
M.~Visser, S.~Weinfurtner, {\og Massive Klein-Gordon equation from a
  Bose-Einstein-condensation-based analogue spacetime\fg}, \emph{Physical
  Review D} \textbf{72} (2005), \cdrnumero 4, p.~044020.

\bibitem{liberati2006analogue}
S.~Liberati, M.~Visser, S.~Weinfurtner, {\og Analogue quantum gravity
  phenomenology from a two-component Bose--Einstein condensate\fg},
  \emph{Classical and Quantum Gravity} \textbf{23} (2006), \cdrnumero 9,
  p.~3129.

\bibitem{butera2017black}
S.~Butera, P.~{\"O}hberg, I.~Carusotto, {\og Black-hole lasing in coherently
  coupled two-component atomic condensates\fg}, \emph{Physical Review A}
  \textbf{96} (2017), \cdrnumero 1, p.~013611.

\bibitem{cominotti2022observation}
R.~Cominotti, A.~Berti, A.~Farolfi, A.~Zenesini, G.~Lamporesi, I.~Carusotto,
  A.~Recati, G.~Ferrari, {\og Observation of Massless and Massive Collective
  Excitations with Faraday Patterns in a Two-Component Superfluid\fg},
  \emph{Physical Review Letters} \textbf{128} (2022), \cdrnumero 21, p.~210401.

\bibitem{finazzi2014entangled}
S.~Finazzi, I.~Carusotto, {\og Entangled phonons in atomic Bose-Einstein
  condensates\fg}, \emph{Physical Review A} \textbf{90} (2014), \cdrnumero 3,
  p.~033607.

\bibitem{kuopanportti2019}
P.~Kuopanportti, S.~Bandyopadhyay, A.~Roy, D.~Angom, {\og Splitting of singly
  and doubly quantized composite vortices in two-component Bose-Einstein
  condensates\fg}, \emph{Phys. Rev. A} \textbf{100} (2019), p.~033615,
  \url{https://link.aps.org/doi/10.1103/PhysRevA.100.033615}.

\bibitem{manni2012dissociation}
F.~Manni, K.~Lagoudakis, T.~Liew, R.~Andr{\'e}, V.~Savona, B.~Deveaud, {\og
  Dissociation dynamics of singly charged vortices into half-quantum vortex
  pairs\fg}, \emph{Nature communications} \textbf{3} (2012), \cdrnumero 1,
  p.~1-7.

\bibitem{seo2015half}
S.~W. Seo, S.~Kang, W.~J. Kwon, Y.-i. Shin, {\og Half-quantum vortices in an
  antiferromagnetic spinor Bose-Einstein condensate\fg}, \emph{Physical Review
  Letters} \textbf{115} (2015), \cdrnumero 1, p.~015301.

\bibitem{stringari_2016}
L.~Pitaevskii, S.~Stringari, \emph{Bose-Einstein Condensation and
  Superfluidity}, International series of monographs on physics, Oxford
  University Press, 2016, \url{https://books.google.it/books?id=\_y4ZswEACAAJ}.

\bibitem{Abad_2013}
M.~Abad, A.~Recati, \emph{The European Physical Journal D} \textbf{67} (2013),
  p.~148, \url{https://doi.org/10.1140/epjd/e2013-40053-2}.

\bibitem{castin2001bose}
Y.~Castin, {\og Bose-Einstein condensates in atomic gases: simple theoretical
  results\fg}, in \emph{Coherent atomic matter waves}, Springer, 2001,
  p.~1-136.

\bibitem{antoine2017efficient}
X.~Antoine, A.~Levitt, Q.~Tang, {\og Efficient spectral computation of the
  stationary states of rotating Bose--Einstein condensates by preconditioned
  nonlinear conjugate gradient methods\fg}, \emph{Journal of Computational
  Physics} \textbf{343} (2017), p.~92-109.

\bibitem{patrick2022origin}
S.~Patrick, A.~Geelmuyden, S.~Erne, C.~F. Barenghi, S.~Weinfurtner, {\og
  {Origin and evolution of the multiply quantized vortex instability}\fg},
  \emph{Physical Review Research} \textbf{4} (2022), \cdrnumero 4, p.~043104.

\bibitem{deNova:PRA2016}
J.~R.~M. de~Nova, S.~Finazzi, I.~Carusotto, {\og Time-dependent study of a
  black-hole laser in a flowing atomic condensate\fg}, \emph{Phys. Rev. A}
  \textbf{94} (2016), p.~043616,
  \url{https://link.aps.org/doi/10.1103/PhysRevA.94.043616}.

\bibitem{Bos16}
P.~Bosch, S.~R. Green, L.~Lehner, {\og Nonlinear Evolution and Final Fate of
  Charged Anti--de Sitter Black Hole Superradiant Instability\fg}, \emph{Phys.
  Rev. Lett.} \textbf{116} (2016), p.~141102,
  \url{https://link.aps.org/doi/10.1103/PhysRevLett.116.141102}.

\bibitem{San16}
N.~Sanchis-Gual, J.~C. Degollado, P.~J. Montero, J.~A. Font, C.~Herdeiro, {\og
  Explosion and Final State of an Unstable Reissner-Nordstr\"om Black Hole\fg},
  \emph{Phys. Rev. Lett.} \textbf{116} (2016), p.~141101,
  \url{https://link.aps.org/doi/10.1103/PhysRevLett.116.141101}.

\bibitem{Eas17}
W.~E. East, F.~Pretorius, {\og Superradiant Instability and Backreaction of
  Massive Vector Fields around Kerr Black Holes\fg}, \emph{Phys. Rev. Lett.}
  \textbf{119} (2017), p.~041101,
  \url{https://link.aps.org/doi/10.1103/PhysRevLett.119.041101}.

\bibitem{Ike19}
T.~Ikeda, R.~Brito, V.~Cardoso, {\og Blasts of Light from Axions\fg},
  \emph{Phys. Rev. Lett.} \textbf{122} (2019), p.~081101,
  \url{https://link.aps.org/doi/10.1103/PhysRevLett.122.081101}.

\bibitem{Madison:PRL2000}
K.~W. Madison, F.~Chevy, W.~Wohlleben, J.~Dalibard, {\og Vortex Formation in a
  Stirred Bose-Einstein Condensate\fg}, \emph{Phys. Rev. Lett.} \textbf{84}
  (2000), p.~806-809,
  \url{https://link.aps.org/doi/10.1103/PhysRevLett.84.806}.

\bibitem{Carusotto:RMP2013}
I.~Carusotto, C.~Ciuti, {\og Quantum fluids of light\fg}, \emph{Rev. Mod.
  Phys.} \textbf{85} (2013), p.~299-366,
  \url{https://link.aps.org/doi/10.1103/RevModPhys.85.299}.

\bibitem{carusotto2008numerical}
I.~Carusotto, S.~Fagnocchi, A.~Recati, R.~Balbinot, A.~Fabbri, {\og Numerical
  observation of Hawking radiation from acoustic black holes in atomic
  Bose--Einstein condensates\fg}, \emph{New Journal of Physics} \textbf{10}
  (2008), \cdrnumero 10, p.~103001.

\bibitem{recati2009bogoliubov}
A.~Recati, N.~Pavloff, I.~Carusotto, {\og Bogoliubov theory of acoustic Hawking
  radiation in Bose-Einstein condensates\fg}, \emph{Physical Review A}
  \textbf{80} (2009), \cdrnumero 4, p.~043603.

\bibitem{wright2009finite}
T.~Wright, A.~Bradley, R.~Ballagh, {\og Finite-temperature dynamics of a single
  vortex in a Bose-Einstein condensate: Equilibrium precession and rotational
  symmetry breaking\fg}, \emph{Physical Review A} \textbf{80} (2009),
  \cdrnumero 5, p.~053624.

\bibitem{jackson2009finite}
B.~Jackson, N.~Proukakis, C.~Barenghi, E.~Zaremba, {\og Finite-temperature
  vortex dynamics in Bose-Einstein condensates\fg}, \emph{Physical Review A}
  \textbf{79} (2009), \cdrnumero 5, p.~053615.

\bibitem{isoshima2003instabilities}
T.~Isoshima, J.~Huhtam{\"a}ki, M.~M. Salomaa, {\og Instabilities of
  off-centered vortices in a Bose-Einstein condensate\fg}, \emph{Physical
  Review A} \textbf{68} (2003), \cdrnumero 3, p.~033611.

\bibitem{fialko2015fate}
O.~Fialko, B.~Opanchuk, A.~Sidorov, P.~Drummond, J.~Brand, {\og Fate of the
  false vacuum: towards realization with ultra-cold atoms\fg}, \emph{EPL
  (Europhysics Letters)} \textbf{110} (2015), \cdrnumero 5, p.~56001.

\bibitem{billam2019simulating}
T.~P. Billam, R.~Gregory, F.~Michel, I.~G. Moss, {\og Simulating seeded vacuum
  decay in a cold atom system\fg}, \emph{Physical Review D} \textbf{100}
  (2019), \cdrnumero 6, p.~065016.

\end{thebibliography}

\end{document}